\documentclass[aps,showpacs,twocolumn,floats,prd,superscriptaddress,nofootinbib]{revtex4-1} 
\usepackage{graphicx,amsmath,amssymb,amstext}
\usepackage{amssymb,amsbsy,amsfonts,amsthm,color}

\usepackage{epsfig}
\usepackage{graphicx}
\usepackage{subfigure}

\graphicspath{{Figures/}}

\begin{document}

\title{ The Entanglement Timescale }

\author{I-Sheng Yang}
\email{isheng.yang@gmail.com}
\affiliation{Canadian Institute of Theoretical Astrophysics, 60 St George St, Toronto, ON M5S 3H8, Canada.}
\affiliation{Perimeter Institute of Theoretical Physics, 31 Caroline Street North, Waterloo, ON N2L 2Y5, Canada.}

\begin{abstract}

We derive the timescale for two initially pure subsystems to become entangled with each other through an arbitrary Hamiltonian that couples them.
The entanglement timescale is inversely proportional to the ``correlated uncertainty'' between the two subsystems, a quantity which we will define and analyze in this paper.
Our result is still applicable when one of the subsystems started in an arbitrarily mixed state, thus it generalizes the well-known ``decoherence timescale'' while coupled to a thermal state.

\end{abstract}

\maketitle

\section{Introduction}

Quantum entanglement is one of the most intriguing properties of quantum mechanics.
From the kinematics alone, one can already derive surprisingly universal results such as the violation of Bell Inequality and the monogamy of entanglement \cite{Bell1964}.
In this paper, we will take a step toward a universal result in the dynamics---the entanglement timescale.

Specifically, we will study two subsystems which started in a pure, product state.
\begin{eqnarray}
|\Psi(t=0)\rangle &=& |\phi\rangle_A \bigotimes |\psi\rangle_B~.
\end{eqnarray}
They then evolve to become entangled with each other through couplings in the Hamiltonian.
\begin{eqnarray}
H &=& \sum_n A_n \bigotimes B_n~.
\label{eq-H}
\end{eqnarray}
Since any Hamiltonian of the full system can be written as a sum of tensor products of subsystem operators, Eq.~(\ref{eq-H}) is completely general.

We will show that there is a well-defined, universal timescale in this problem.
\begin{equation}
T_{ent}^{-2} = 
\sum_{m,n} \left( \langle A_mA_n\rangle - \langle A_m\rangle\langle A_n\rangle \right)
\left( \langle B_mB_n\rangle - \langle B_m\rangle\langle B_n\rangle \right)~.
\label{eq-result}
\end{equation}
This indicates how fast the two subsystems become entangled with each other.

The physical meaning of Eq.~(\ref{eq-result}) is more obvious if we consider a simpler system.
\begin{equation}
H_{tot} = H_A + H_B + O_AO_B~.
\label{eq-Hmin}
\end{equation}
Here, the two subsystems are minimally coupled.
Eq.~(\ref{eq-result}) then gives a simple answer.
\begin{equation}
T_{ent} = (\Delta O_A \Delta O_B)^{-1}~.
\label{eq-Tmin}
\end{equation}
Here $\Delta O_A$ and $\Delta O_B$ are the standard definition of the quantum uncertainties of the corresponding observables in the initial state.
Thus, the simple and important lesson here is that {\bf entanglement between two subsystems are due to the quantum uncertainty of their coupling.} 
The quantity inside the square-root of Eq.~(\ref{eq-result}) can be understood as a nontrivial generalization of quantum uncertainty.
We will call it {\bf correlated quantum uncertainty} and explain a few of its properties in Sec.\ref{sec-CQU}.

The dynamics of entanglement has always been an active research topic.
It is closely related to the famous decoherence problem of quantum mechanics \cite{Zurek03}.
A pure state will demonstrate a lot of intriguing quantum properties.
Only an entangled subsystem will behave manifestly classically.
Since most of the observations we make in this world are consistent with classical behaviors, it is important to understand how most of the subsystems become entangled with each other. 
Some even argued that this is closely related to consciousness and free-will \cite{Teg14}.
A natural challenge in such study is that the full system is absolutely macroscopic.
It is often impractical or impossible to explicitly model it.
A well-known result is the decoherence timescale when coupled to a thermal environment \cite{Zurek86}.
That can be viewed as a special case of the general entanglement timescale we derived here.
Hopefully, our more general result it can help to shed some new light on this long-standing topic.

A more modern motivation to study the dynamics of entanglement is quantum computing.
One would like a subsystem to stay not entangle with the environment as long as possible to perform a quantum computation \cite{ChuLaf95}.
Our result may help to derive a universal bound on how well can that be done.

Our main motivation is to study the unitarity problem for quantum field theory (QFT) in curved spacetime.
It is widely believed that the true theory of quantum gravity exists, and the geometry can ultimately be described by some wave-function formalism.
Therefore, it is by-definition possible for the wave-function of particles in QFT to become entangled with the geometry.
Unfortunately, without an actually well-established theory of quantum gravity, it is difficult to study such entanglement.
The hope is that since our result does not rely on any specific assumptions of the subsystems, it may eventually provide a way to circumvent this obstacle.
One might be able to still learn something about the unitarity loss of QFT without knowing the exact theory of quantum gravity.

The rest of the paper goes like the following.
In Sec.\ref{sec-proof}, we will present the derivation of Eq.~(\ref{eq-result}).
In Sec.\ref{sec-CQU}, we discuss the properties of correlated quantum uncertainty.
In Sec.\ref{sec-dis}, we will discuss the general implication and future directions to further generalize our result.

\section{The Entanglement Timescale}
\label{sec-proof}

\subsection{The Entanglement Measure}
\label{sec-EM}

As stated in the introduction, we will consider a general Hamiltonian and a pure product state.
\begin{eqnarray}
H &=&\sum_n H_n =  \sum_n A_n \bigotimes B_n~, \\
\rho(0) &=& \rho_A(0) \bigotimes \rho_B(0)~.
 \label{eq-init}
\end{eqnarray}
\footnote{The total Hamiltonian has to be Hermitian, but the individual subsystem operators $A_n$ and $B_n$ need not be.
}
Being a pure state, we have ${\rm tr}_A (\rho_A^2)={\rm tr}_A (\rho_A)=1$ in the beginning.
As the system evolves, its {\bf purity}, ${\rm tr}_A (\rho_A^2)$ will start to decrease.
We will monitor the entanglement by keeping track of how the value of its purity evolves.

There are other quantities which can also keep track of entanglement, such as the von Neumann entropy.
We chose to monitor purity for a few reasons.
First of all, it is mathematically simple.

The more physical reason is that knowing how purity deviates from 1 actually teaches us a lot more implicit lessons.
Let us write down the eigenvalues of $\rho_A$ as
\begin{eqnarray}
\lambda_1 = (1-\epsilon)~, \ \ \ 
\lambda_{i>1} = \epsilon_i~, \ \ \ 
\sum_{i=2}^N \epsilon_i = \epsilon~,
\end{eqnarray}
where $N$ is the Hilbert space dimension of subsystem $A$.
When all $\epsilon_i$'s are small, the dynamics of purity is controlled by how $\epsilon$ grows with time.
In fact, all Renyi-n entropies with integer $n\geq2$ are dominated by the same behaviour of $\epsilon$.
Thus studying purity is already covering a large portion of the full entanglement spectrum.

In fact, this is also sufficient to provide bounds on the entanglement entropy, which depends on the Hilbert space dimension $N$.
\begin{eqnarray}
S_A &\geq& -(1-\epsilon)\ln(1-\epsilon) - \epsilon\ln\epsilon~, \\
S_A &\leq& -(1-\epsilon)\ln(1-\epsilon) - \epsilon\ln\epsilon + \epsilon\ln(N-1)~.
\end{eqnarray}
As we can see, when $\epsilon<N^{-1}$, this is a relatively small range.
Thus knowing the initial dynamics of $\epsilon$ also tells us almost everything about entanglement entropy, at least for a while.

\subsection{Dynamics}

\begin{eqnarray}
& & \frac{d}{dt} {\rm tr}_A (\rho_A^2)\bigg|_{t=0} 
\label{eq-dr2} \\ \nonumber
&=&
i ~ {\rm tr}_A \bigg[ {\rm tr}_B(H\rho-\rho H) ~ {\rm tr}_B\rho 
+  {\rm tr}_B\rho ~ {\rm tr}_B(H\rho-\rho H)\bigg]
\\ \nonumber &=&
i~\sum_n {\rm tr}_B (B_n\rho_B) 
\bigg[ {\rm tr}_A(A_n\rho_A^2) - {\rm tr}_A(\rho_A^2 A_n) \bigg]=0~.
\end{eqnarray}
It is straightforward to see that as long as the total density matrix takes the product form, the time derivative is always zero.
That is because when the two subsystems are not entangled, ${\rm tr}_A (\rho_A^2)$ is already at its maximal value allowed by the dynamics.

Thus, starting from an unentangled system, the nontrivial dynamics of entanglement comes from the second derivative.
\begin{eqnarray}
& & \frac{d^2}{dt^2}{\rm tr}_A (\rho_A^2)\bigg|_{t=0} 
 \\ \nonumber
&=&
-{\rm tr}_A \bigg\{
2\bigg[ {\rm tr}_B(H\rho-\rho H) \bigg]^2
\\ \nonumber
& & \ \ \ \ \ \ \ \ \ 
+{\rm tr}_B\rho~{\rm tr}_B(H^2\rho -2H\rho H+\rho H^2)
\\ \nonumber
& & \ \ \ \ \ \ \ \ \ 
+{\rm tr}_B(H^2\rho -2H\rho H+\rho H^2)~{\rm tr}_B\rho
\bigg\} 
\\ \nonumber
&=& -4\sum_{m,n}
\bigg(
{\rm tr}_A(\rho_A^2 A_mA_n) 
- {\rm tr}_A(A_m \rho_A A_n \rho_A) 
\bigg)
\\ \nonumber & & \ \ \ \ \ \ \ \ \ 
\bigg(
{\rm tr}_B (B_m B_n \rho_B ) - {\rm tr}_B (B_m\rho_B)~{\rm tr}_B (B_n\rho_B)
\bigg)~.
\end{eqnarray}

When the subsystem $A$ starts in a pure state, we can use two properties to simplify the above result.
First of all, that allows us to use $\rho_A^2=\rho_A$.
Secondly, we can go into the basis that $\rho_A^{ij} = \delta_{1i}\delta_{1j}$.
In this basis, it becomes obvious that
\begin{equation}
{\rm tr}_A(A_m\rho_AA_n\rho_A) = A_m^{11}A_n^{11} =
 {\rm tr}_A(A_n\rho_A)~{\rm tr}_A(A_m\rho_A)~.
 \label{eq-prod}
\end{equation}
Therefore, the second derivative takes a symmetric, universal form.
\begin{eqnarray}
& & \frac{d^2}{dt^2}{\rm tr}_A (\rho_A^2) \bigg|_{t=0} =
\label{eq-final} \\ \nonumber
 &-4& \sum_{m,n} \bigg( \langle A_mA_n\rangle - \langle A_m\rangle\langle A_n\rangle \bigg)
\bigg( \langle B_mB_n\rangle - \langle B_m\rangle\langle B_n\rangle \bigg)~.
\end{eqnarray}
This implies that the evolution near $t=0$ is given by
\begin{equation}
{\rm tr}_A\rho_A^2 \approx 1-2\epsilon \approx 1 - 2\frac{t^2}{T_{ent}^2} + ......~,
\label{eq-time}
\end{equation}
with the entanglement timescale $T_{ent}$ defined in Eq.~(\ref{eq-result}).

\section{Correlated Quantum Uncertainty}
\label{sec-CQU}

Let us take a closer look at the R.H.S. of Eq.~(\ref{eq-final}).
We will define the summed quantity as {\bf correlated quantum uncertainty} between the two subsystems.
\begin{equation}
\Delta_{AB}^2 = \sum_{m,n} \bigg( \langle A_mA_n\rangle - \langle A_m\rangle\langle A_n\rangle \bigg)
\bigg( \langle B_mB_n\rangle - \langle B_m\rangle\langle B_n\rangle \bigg)~.
\label{eq-CQU1}
\end{equation}
Such a name is motivated by its several properties.
First of all, all terms with $m=n$ are directly related to the standard definition of quantum uncertainties of the subsystem operators.
\begin{equation}
\bigg( \langle A_n^2\rangle - \langle A_n\rangle^2 \bigg)
\bigg( \langle B_n^2\rangle - \langle B_n\rangle^2 \bigg)
= (\Delta A_n)^2 (\Delta B_n)^2~.
\label{eq-unc}
\end{equation}
Note that this is not equal to the quantum uncertainty of the coupling term $A_n\bigotimes B_n$.
One obvious difference is that as long as either one of the subsystem operator has zero uncertainty, Eq.~(\ref{eq-unc}) vanishes, but the uncertainty of the entire coupling term does not have to vanish.
Likewise, the correlated quantum uncertainty is also not just the quantum uncertainty of the total Hamiltonian.
The summation over $m$ and $n$ shows that it depends crucially on the subsystem separation.

Another obvious property is that, as long as one subsystem operator is proportional to the identity operator, then such term vanishes in the correlated quantum uncertainty.
\begin{equation}
\langle A_m I \rangle - \langle A_m\rangle\langle I \rangle = 0~.
\label{eq-van}
\end{equation}
That means the ``self-Hamiltonian'' of either subsystems does not contribute to the correlated uncertainty.
Only a coupling term, in which both $A_n$ and $B_n$ are nontrivial, will contribute.
This also means that if one adds a constant to any subsystem operator, for example $A_n \rightarrow (A_n + a)$, the correlated uncertainty does not change.
Combine the two properties above, we get the entanglement timescale for minimally coupled subsystems as stated in the introduction, Eq.~(\ref{eq-Tmin}).

We know that the second derivative of ${\rm tr}_A(\rho_A^2)$ must be negative, since its value is already at maximum.
Such fact is not obvious from the last line of Eq.~(\ref{eq-final}).
Although the diagonal terms with $m=n$ must be positive, the cross terms can be negative.
Here we will quickly prove that the entire sum is indeed positive definite.
\begin{eqnarray}
\nonumber & & \sum_{m,n}\bigg( \langle A_mA_n\rangle - \langle A_m\rangle\langle A_n\rangle \bigg)
\bigg( \langle B_mB_n\rangle - \langle B_m\rangle\langle B_n\rangle \bigg)
\\ \nonumber &=&
\sum_{m,n} 
\bigg( \sum_i  A_m^{1i}  A_n^{i1} - A_m^{11} A_n^{11} \bigg)
\bigg( \sum_j  B_m^{1j}  B_n^{j1} - B_m^{11} B_n^{11} \bigg)
\\ \nonumber &=&
\sum_{i,j}^{\neq1} \left(\sum_m A_m^{1i}B_m^{1j} \right)
\left(\sum_n A_n^{i1}B_n^{j1} \right)
\\ \label{eq-positive} &=&
\sum_{i,j}^{\neq1} \left|\sum_m A_m^{1i}B_m^{1j} \right|^2 \geq0~.
\end{eqnarray}
\footnote{
In the last line, we seemed to have used the property that these matrices are Hermitian.
Actually they do not have to be.
Without loss of generality, we can choose to rewrite the product-sum Hamiltonian in pairs with $A_{2n} = A_{2n+1}^\dagger$ and $B_{2n} = B_{2n+1}^\dagger$.
It is then just a simple relabeling of dummy variables.
}
We have simply expanded in the basis that $\rho_A^{ij}=\rho_B^{ij} = \delta_{1i}\delta_{1j}$.
\footnote{The two subsystems may not have the same Hilbert-space dimensions. 
However, when they are in pure states, we can still put both of them into the same simple form.}
Thus, we conclude that for pure, product states, the correlated quantum uncertainty is positive definite, just like the standard quantum uncertainty.

We can further rewrite the last line of Eq.~(\ref{eq-positive}) into the following form.
\begin{eqnarray}
\Delta_{AB}^2 &=&
\label{eq-CQU2}
\sum_{i,j}^{\neq1} \left|\sum_m A_m^{1i}B_m^{1j} \right|^2
= \sum_{i,j}^{\neq1} \left| H^{(1i)(1j)} \right|^2
\\ \nonumber
&=& \sum_{i,j} H^{(1i)(1j)}H^{(i1)(j1)} - \sum_i H^{(1i)(11)}H^{(i1)(11)}
\\ \nonumber
&-& \sum_j H^{(11)(1j)}H^{(11)(j1)} + (H^{(11)(11)})^2
\\ \nonumber
&=& \langle H^2\rangle_{AB} - \langle \langle H\rangle_B^2 \rangle_A
- \langle \langle H\rangle_A^2\rangle_B + \langle H\rangle_{AB}^2~.
\end{eqnarray}
This highlights the role of subsystem separation in the definition of correlated quantum uncertainty.

\section{Discussion}
\label{sec-dis}

We should remind the readers that the full dynamics of entanglement can be very complicated.
The only reason why we get a universal behavior is that we started from a rare and the simplest situation---a pure, product state.
Entanglement will grow as we predicted initially, but higher-order terms in Eq.~(\ref{eq-time}) can become relevant anytime, even before $T_{ent}$.

For example, we can choose an initial state such that the correlated quantum uncertainty vanishes, $\Delta_{AB}=0$.
Applying our result na\"ively, we have $T_{ent}=\infty$.
It simply means that the higher order terms in Eq.~(\ref{eq-time}) becomes relevant right away.
Another calculation is required to understand their dynamics.
At very least, higher order terms should include the contribution from $[H,\Delta_{AB}]$.
It tells us that $\Delta_{AB}$ can first evolve into nonzero values, which in turn allows the two subsystems to become entangled. 

Even when $T_{ent}$ is finite, it is in-principle possible to have higher order terms to kick in before $T_{ent}$, and even conspire to prohibit the growth of entanglement.
Nevertheless, we think such arrangement, if indeed possible, is highly contrived.
The fully nonlinear evolution of entanglement has been studied in several examples with minimal couplings.
They have all shown a clear agreement with our result.
Entanglement basically approaches maximum in a timescale inversely proportional to the quantum uncertainty \cite{PikZyc15, AdlBas16, Yan17}.
Thus, we advocate that two coupled subsystems generically will become significantly entangled at the entanglement timescale.

There is another interesting direction to generalize our result.
During our derivation in Sec.\ref{sec-proof}, we actually did not use the fact that the subsystem states are pure before Eq.~(\ref{eq-prod}).
In fact, we only needed to assume that subsystem $A$ is in a pure state to reach Eq.~(\ref{eq-CQU1}).
Thus, Eq.~(\ref{eq-CQU1}) is valid even if subsystem $B$ is in any mixed state.
Thus, our result is also applicable to the well-known case of decoherence due to a thermal environment \cite{Zurek86}.

One might worry that in such generalizations, the loss of purity can be due to either entanglement, or just the in-flow of non-purity from the other subsystem. 
Thus the timescale can only be understood as the decoherence timescale instead.
Such point of view is not incorrect, but we provide an alternative picture to circumvent that problem.
When systems $A$ and $B$ are generally mixed, but not correlated or entangled with each other, one can introduce two hidden systems: $A'$ which purifies $A$, and $B'$ which purifies $B$.
Our same calculation then applies to the entanglement between $(A'A)$ and $(B'B)$ without ambiguity.
Since the hidden systems has no dynamics, it is natural that the entanglement timescale depends only on the states of $A$ and $B$. 
In fact, the two forms of correlated quantum uncertainty, Eq.~(\ref{eq-CQU1}) and (\ref{eq-CQU2}), both are well-defined and calculable on mixed subsystems.

Finally, we should point out that our result sounds a serious warning call to the unitarity of QFT in curved spacetime.
If the theory of quantum gravity exists, then QFT in curved spacetime is a semi-classical approximation.
The geometry is a subsystem that we only have classical descriptions.
Quantum uncertainty plays no specific role in a classical description.
However, as we have shown, quantum uncertainty is what controls entanglement.
Thus, within the semi-classical framework, it is impossible to guarantee that QFT does not become entangled with geometry and loses its subsystem unitarity.
Such issue was already discussed by examples in \cite{Unr12, Yan17}.
Armed with the simple result here, we will come back to the unitarity problem of QFT in a future paper \cite{TBA}.

\acknowledgments

We thank Henrique Gomes for inspiring discussions. 
We also thank Jesse Cresswell, Ben Freivogel, Ted Jacobson, Dimitrios Krommydas, Jess Riedel and Wojciech Zurek for helpful comments.
This work is supported by the Canadian Government through the Canadian Institute for Advance Research and Industry Canada, and by Province of Ontario through the Ministry of Research and Innovation.

\appendix

\bibliography{all_active}

\appendix

\end{document}